\documentclass[journal]{IEEEtran}
\usepackage{color,soul}

\ifCLASSINFOpdf
   \usepackage[pdftex]{graphicx}
\else
\fi
\usepackage{amsmath}
\usepackage{amsfonts}

\usepackage{subfiles}
\ifCLASSOPTIONcompsoc
  \usepackage[caption=false,font=normalsize,labelfont=sf,textfont=sf]{subfig}
\else
  \usepackage[caption=false,font=footnotesize]{subfig}
\fi
\begin{document}
%
\title{Distributed Kuramoto Self-Synchronization of Vehicle Speed Trajectories in Traffic Networks}
%
%
%

\author{Manuel~Rodriguez,
        Hosam~Fathy
\thanks{
}
\thanks{
}
\thanks{Manuscript received April , 2019; revised August , 2019.}}

\maketitle

\begin{abstract}
This paper presents a distributed synchronization strategy for connected and automated vehicles in traffic networks. The strategy considers vehicles traveling from one intersection to the next as waves. The phase angle and frequency of each wave map to its position and velocity, respectively. The goal is to synchronize traffic such that intersecting traffic waves are out of phase at every intersection. This ensures the safe collective navigation of intersections. Vehicles share their phase angles through the V2X infrastructure, and synchronize these angles using the Kuramoto equation. This is a classical model for the self-synchronization of coupled oscillators. The mapping between phase and location for vehicles on different roads is designed such that Kuramoto synchronization ensures safe intersection navigation. Each vehicle uses a constrained optimal control policy to achieve its desired target Kuramoto phase at the upcoming intersection. The overall outcome is a distributed traffic synchronization algorithm that simultaneously tackles two challenges traditionally addressed independently, namely: coordinating crossing at an individual intersection, and harmonizing traffic flow between adjacent intersections. Simulation studies highlight the positive impact of this strategy on fuel consumption and traffic delay time, compared to a network with traditional traffic light timing. 

\end{abstract}


%
\IEEEpeerreviewmaketitle

\section{Introduction}

\IEEEPARstart{A}{utonomous} and connected vehicles are expected to improve the efficiency of road networks by increasing throughput, reducing fuel consumption and reducing travel time. Optimistic forecasts anticipate up to 45\% reduction in fuel consumption through the implementation of these technologies \cite{Chen2019}. Furthermore, an important portion of these savings is expected to come from the automation of traffic intersections.


Coordinating traffic at intersections requires solving two different problems, at two different scales. The first is servicing conflicting flows at an intersection so that vehicles do not collide with each other; in other words, deciding who crosses when. We refer to this problem as the intra-junction coordination problem.  The second problem is harmonizing the flow between adjacent intersections to reduce the amount of energy vehicles waste due to frequent acceleration and braking; we refer to this problem as the inter-junction coordination problem. Most of the literature on intersection control focuses on one of these problems individually. Some approaches combine separate solutions, and evaluate their performance when combined.

\subsection {Literature review}
Approaches in the literature differ mostly in their assumptions of which agents communicate wirelessly with each other, and which agents are being controlled by the proposed strategy. Agents are essentially of two types: vehicles, which can be connected and autonomous or human-driven, and coordinators, which can be traditional traffic lights or computers that communicate wirelessly with the incoming traffic. In the following, we review work that considers coordination between traffic lights, coordination between CAVs and upcoming traffic lights, coordination between CAVs themselves, and finally coordination between CAVs and centralized controllers, which in turn coordinate with each other. 

First, we consider coordination between traffic lights that can exchange signal phase and timing information with each other. Traffic networks currently solve the intra-junction problem through traffic signalization. When these traffic signals are actuated (i.e traffic lights), proper tuning of signal timing and offset can also tackle the inter-junction problem. Indeed, an early impact of connectivity in improving traffic has come from coordinating traffic lights so that their offsets give rise to ``green waves" where vehicles encounter green lights in sequence, effectively avoiding start-stop behaviour. Currently, through loop detection, traffic lights can adapt their timing using strategies such as SCOOT \cite{Hunt1981} and SCAAT \cite{Lowrie1982}. Other approaches in the literature are based on optimal control \cite{ABOUDOLAS2010680}, fuzzy logic \cite{TRABIA1999353}, machine learning \cite{JIN2018236,ABDOOS20131575}, and game theory \cite{Vasirani2012}. 

Of specific relevance to the work presented in this paper, several proposed traffic light control approaches make use of the Kuramoto equation for self-synchronizing oscillators. Sekiyama et al. proposed such an approach, where they used Kuramoto synchronization to adjust signal phase and timing \cite{Sekiyama2001}. 
This work was further expanded in \cite{Nishikawa2004,Lammer2006,Donner2009}. In general the problem can be thought of as material transport problem in a directed graph, as explained by Lammer et. al. in \cite{Lammer2006}. 

As cars themselves become connected and autonomous, they can coordinate with each other and with the traffic lights. Several algorithms are explored using both of these communication configurations. For example, assuming vehicle-to-infrastructure communication, the inter-junction problem can be addressed through predictive trajectory planning. Specifically, vehicles that receive signal phase and timing information from traffic lights can plan their trajectories to avoid stopping at red lights. This approach has been studied in \cite{Mandava2009,He2015,HomChaudhuri2016,Du2017,Wan2016,Rodriguez2018}, and has shown promising fuel economy and throughput improvements. The idea consists of translating signal phase and timing (SPAT) information into constraints for a model-predictive trajectory optimization problem that seeks to reduce fuel consumption. This body of work further exemplifies the importance of considering the intersection management problem in both of its scales.  For example, in \cite{Rodriguez2018} we show how even with the use of optimization to generate trajectories in an arterial corridor, improper timing of the traffic lights can halve the benefits in fuel consumption.

Assuming vehicle fleet connectivity, some approaches avoid the use of traffic lights altogether by having the vehicles coordinate crossing times with each other, or with a centralized coordinator. For comprehensive reviews of this literature, the reader is pointed towards references \cite{Rios-torres2016}, and \cite{Chen2016}. In these two reviews, the different approaches are classified by whether decisions are made in a centralized or decentralized manner, whether they are the product of heuristic rules or the solution to an optimization problem (\cite{Tallapragada2019,Rios-torres2017}), or whether they are reservation-based approaches (\cite{Dresner2008a,Hausknecht2011}) versus trajectory planning approaches (\cite{Makarem2013,Kim2015}). In general the problem is solved in two layers. The first one determines vehicles' crossing times or sequence, while the second controls the vehicles' speeds to achieve the agreed-upon crossing time. The approach proposed in this work follows a similar structure. Other approaches based on the formation of virtual platoons are also of particular relevance to this work, because they make use of multi-agent consensus strategies, which can be thought of as linear counter-parts of the Kuramoto equation. Vaio et al. \cite{Vaio2019} propose a decentralized protocol that projects vehicles in conflicting roads into the same coordinate system, namely a distance to the upcoming intersection. Through a heuristic algorithm vehicles negotiate desired inter-vehicular distances, and they use modified consensus to achieve the desired formation. The method is evaluated for a single intersection.

Finally, we consider approaches that have both connected centralized agents, that can communicate with each other at different intersections, and connected vehicles that exchange information with these coordinators and among themselves. Recent studies explore how these approaches can attempt to solve both the inter-junction and intra-junction problem. In \cite{Ashtiani2018}, a centralized reservation-based controller at each intersection communicates its decisions to both the vehicles it is in charge of scheduling and to the controllers at neighboring intersections. The crossing time decisions are made by solving a mixed-integer linear program that considers the information it receives from its adjacent intersection managers. The approach is evaluated both with and without coordination between intersection managers, showing that when intersection controllers can communicate with each other, fuel consumption benefits double. A similar approach is taken in \cite{Wuthishuwong2017}, where connected centralized schedulers take into account each other's information when making reservation decisions. In this case, coordination between the schedulers is achieved by the use of multi-agent consensus, as opposed to optimization.

\subsection{Paper Contribution}

The above literature highlights the breadth of different approaches that are proposed to solve the autonomous intersection management problem. 
From this literature, we identify the following key lessons.
First, Kuramoto models and other consensus-based approaches have been successfully used to coordinate traffic lights and centralized intersection coordinators, but they have not been explored as means to coordinate autonomous vehicles themselves. Second, approaches that consider multiple intersections and the coupling between them can yield larger fuel savings compared to localized controls. However, most approaches that solve both intra- and inter-junction problems rely on some sort of centralized agent that couples intersections. This paper proposes an approach to solve the autonomous intersection management problem at both levels using the non-linear consensus equation known as the Kuramoto model. The use of Kuramoto allows vehicles to first agree upon the current state of the intersections (i.e. which flow is being serviced), and then to synchronize with the intersections. The conference version of this paper \cite{Rodriguez2019} introduced the basic idea underlying this strategy; namely, the fact that through mapping phase and frequency to position and velocity and synchronizing using Kuramoto, vehicles can agree on crossing times that solve both the intra- and inter-junction problem. In addition, this archival version of the work  adds the following key contributions. First, we pair the synchronizing Kuramoto layer with a more sophisticated optimization-based tracker, as opposed to the back-stepping tracker used in \cite{Rodriguez2019}. Second, we now allow right turns in the network, which puts us closer to adapting the strategy to more realistic traffic scenarios. Third, we evaluate the impact of our  strategy on fuel consumption and delay time as compared to human-driven vehicles in the presence of traffic lights. The fuel savings are then explained by correlating them to to changes in vehicle behavior (e.g., reductions in energy loss due to the braking portions of start-stop driving).
\section{Proposed Strategy}
We consider a grid of interconnected intersections in an urban traffic network. We assume that all incoming vehicles are autonomous, capable of vehicle-to-vehicle (V2V) communication, and can interact with all the vehicles in the network (i.e. all-to-all connectivity). Less restrictive communication topologies are possible as shown in the Kuramoto consensus literature \cite{Jadbabaie2004}, and do not alter the fundamental ideas behind this work.

The idea behind our proposed strategy consists of mapping the position and velocity of each vehicle to a corresponding virtual phase and frequency. The vehicles exchange phase information through V2V communication, and compute the dynamics of their phases using the Kuramoto equation. This naturally drives them to synchronize. From the phase trajectories, vehicles determine the times and velocities at which they need to arrive to the upcoming intersections. With this information, they formulate a linear quadratic optimal control problem that can be solved at each time step to determine the acceleration command that will place them at the intersection at the right time, with the right speed. The mapping between phase and position must satisfy certain constraints for this strategy to produce the desired behaviour, that is, safe crossing at intersections (which solves the intra-junction problem) and smooth crossing between intersections (which addresses the inter-junction problem). 

In this section we describe in detail how the proposed strategy can be implemented in a network of roads with or without right turns, where all vehicles are autonomous and inter-connected. 

\subsection{Kuramoto Synchronization}

The literature on traffic light synchronization using the Kuramoto equation works by describing the agents (i.e the traffic lights) as oscillators and establishing a mapping between the phase of the agent and a control action (i.e switching from green to yellow, or red). In our proposed strategy, where the vehicles are the agents as opposed to the traffic lights, the mapping relates the phase of the vehicles to a position along the road. For each road segment $p$, we define a mapping $g_p(\theta)$ that relates the phase $\theta$ of a vehicle to the vehicle's desired distance to the intersection along the curvature of the road $s^d$:

\begin{equation} \label{gr}
s^d = g_p(\theta)
\end{equation}

We choose $g_p$ to be an affine function of phase; it can therefore be described by two parameters. We call these parameters the wavelength $\lambda$ and the offset $\phi$, where $\lambda$ is the slope and $\phi$ the zero crossing. For a given road $p$, the mapping is then:

\begin{equation} \label{grDef}
g_p(\theta) = (\theta-\phi_{p})\frac{\lambda_{p}}{2\pi}
\end{equation}

\noindent We can think of this map as having wrapped the length along the road around a circle of radius $\frac{\lambda_p}{2\pi}$, and rotated it by an angle $\phi_p$

Assuming steady state tracking of the desired distance to the intersection $s^d$, it follows from the definition of $g_p$ that a vehicle $i$ on road $p$ will reach the intersection when its phase $\theta_i$ is equal to the corresponding offset $\phi_p$. It also follows that two vehicles on road $p$ with a phase difference of some multiple $k$ of $2\pi$, will be separated by $k\lambda_p$ meters. Mathematically, these two properties of our mapping can be expressed as:

\begin{equation} \label{gr0}
g_p(\phi_p) = 0
\end{equation}

\begin{equation} \label{grLambda}
g_p(\theta+2k\pi) - g_p(\theta) = k\lambda_p
\end{equation}

We have yet to define one of the main descriptors of an oscillator: its natural frequency. Since phase is mapped onto position, frequency will be mapped onto velocity. Indeed, from differentiation in time of Eq. \eqref{grDef}, we have a definition of desired vehicle velocity:

\begin{equation} \label{vdDef}
v^d = \dot{\theta}\frac{\lambda_p}{2\pi}
\end{equation}

Under this definition, it follows that the natural frequency $\omega$ of a vehicle is simply the frequency corresponding to the constant nominal desired velocity the vehicle would like to travel at. In our proposed strategy a key constraint is that all vehicles have the same natural frequency $\omega_n$. As such, a road segment $p$ is characterized not only by its wavelength $\lambda_p$, but also by a nominal speed $v_{n,p}$, such that the following constraint is always satisfied\footnote{The possibility of allowing multiple nominal speeds on multi-lane road segments is not precluded by this problem formulation, since the different lanes can correspond to different wavelengths.}:

\begin{equation} \label{OmegaConstraint}
\omega_n = 2\pi \frac{v_{n,p}}{\lambda_p} 
\end{equation}

Now that we have a definition of phase and frequency as they relate to desired position and velocity, we consider the dynamics of this phase variable. Specifically, we impose that these dynamics be governed by the Kuramoto equation. This equation was introduced in 1975 to model the dynamics of populations of weakly coupled oscillators that exhibit self-synchronizing behaviour. Synchronization refers to oscillators with different natural frequencies influencing each other to oscillate at the same frequency and a constant phase difference. This occurs mostly in biological systems like populations of flashing fireflies. The governing equation, as proposed by Kuramoto in \cite{Kuramoto1975}, is as follows:

\begin{equation} \label{KuramotoFull}
\dot{\theta_i}(t) = \omega_{i} + \frac{1}{N}\sum^{N}_{j=1} K_{ij}\sin(\theta_j(t)-\theta_i(t)) 
\end{equation}

In this formulation, the instantaneous frequency of oscillation $\dot{\theta_i}$ is given by the oscillator's natural frequency $\omega_i$ plus the coupling term to all other oscillators based on the sine of their difference in phase multiplied by a coupling term $K_{ij}$.

For all-to-all symmetric coupling, that is $K_{ij}=K$, and a monotonic and uni-modal distribution of natural frequencies $p(\omega)$, the behaviour and stability of the system is well-understood \cite{Strogatz2000}. To illustrate this behaviour, it is useful to express the model in its mean-field form, by introducing the order parameter:

\begin{equation} \label{r}
r(t) e^{\Psi(t) i} = \frac{1}{N}\sum^{N}_{j=1} e^{\theta_j(t) i}
\end{equation}

\begin{figure}[h]
    \centering
    \includegraphics[]{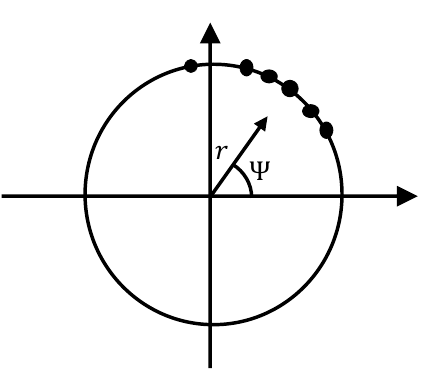}
    \caption{The order parameter has magnitude $r$, the coherence, and phase $\Psi$, the mean phase}
    \label{unitCircle}
\end{figure}

If each oscillator is thought of as a particle orbiting around the unit circle, the order parameter is the centroid of all oscillators, as shown in Fig. \ref{unitCircle}. The Kuramoto equation can then be rearranged in terms of $r$ and $\Psi$:

\begin{equation} \label{KuramotoMean}
\dot{\theta_i}(t) = \omega_{i} + r(t) K \sin(\Psi(t)-\theta_i(t)) 
\end{equation}

In this form, one can see that the $i^{th}$ oscillator is pulled towards the mean phase $\Psi$ with an effective coupling $Kr$. The coherence $r$ takes values from 0 to 1, where 0 represents all oscillators orbiting incoherently and 1 represents all of them sharing the same phase.

When the coupling between oscillators is $K=0$, agents orbit the unit circle in complete incoherence and the value of $r$ fluctuates around $0$. As the coupling strength is increased, incoherent behaviour persists until a critical coupling threshold $K_c$ is exceeded. For these larger values of $K$, a subset of oscillators synchronize and start recruiting more and more oscillators. Indeed, a positive relationship exists between the coherence $r$ and the coupling strength $Kr$. From Eq. \eqref{KuramotoMean} we can see that the stronger the coupling, the more the oscillator is pulled towards the mean phase, and as more oscillators orbit near the mean phase, the coherence $r$ increases. Finally, the value of $r$ saturates at some final value below, but near 1, around which it fluctuates.

For normal distributions of natural frequencies and large enough coupling, the resulting behaviour corresponds to all oscillators orbiting with the mean frequency of the original distribution (this is called frequency entrainment) and maintaining a constant phase difference between each other (this is called phase-locking). In the particular case of all natural frequencies being the same, all vehicles phase lock to the mean phase exactly, with no constant phase difference between them, and $r$ converges to 1 exactly. Using the frequency given in Eq. \eqref{OmegaConstraint}, we can write the dynamics of $\theta_i$ as:

\begin{equation} \label{KuramotoMeanWn}
\dot{\theta_i}(t) = \omega_{n} + r(t) K \sin(\Psi_i(t)-\theta_i(t)) 
\end{equation}

\noindent where the local mean phase $\Psi_i$ is the closest projection to $\theta_i$ of the overall mean. That is:

\begin{equation} \label{MeanPhaseProjection}
\begin{gathered}
\Psi_i(t) = \min_k \{\Psi(t)+2k\pi \}\\
\text{subject to:}\\
-\pi \leq ||\Psi(t)+2k\pi - \theta_i(t)|| \leq \pi\\
k\in \mathbb{Z}
\end{gathered}
\end{equation}

By collapsing the distribution of natural frequencies of the oscillators into a single point (i.e $p(\omega_n)=1$), we force all the phases of the system to converge to the mean phase plus some multiple of $2\pi$, or, in other words, to its closest mean phase $\Psi_i(t)$.

\begin{figure}[t]
\centering
\subfloat[Phase and Mean phase trajectories]{\includegraphics[width=\columnwidth]{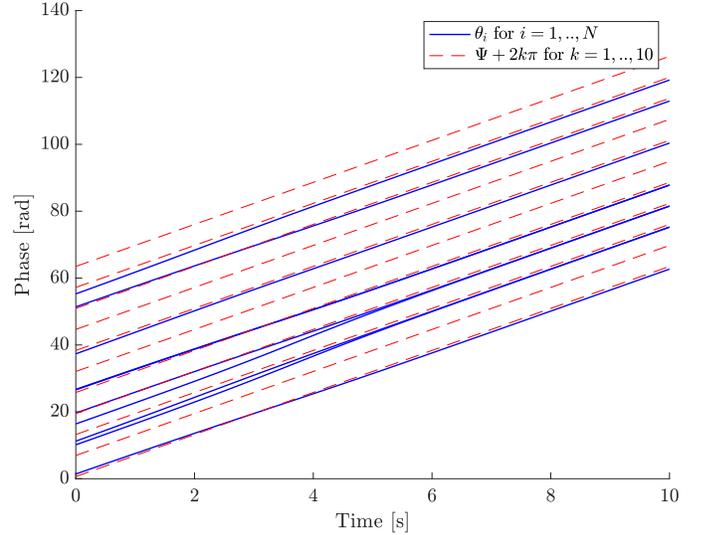}%
\label{phaseAndMeanPhase}}
\hfil
\subfloat[Evolution of frequencies $\theta_i$]{\includegraphics[width=\columnwidth]{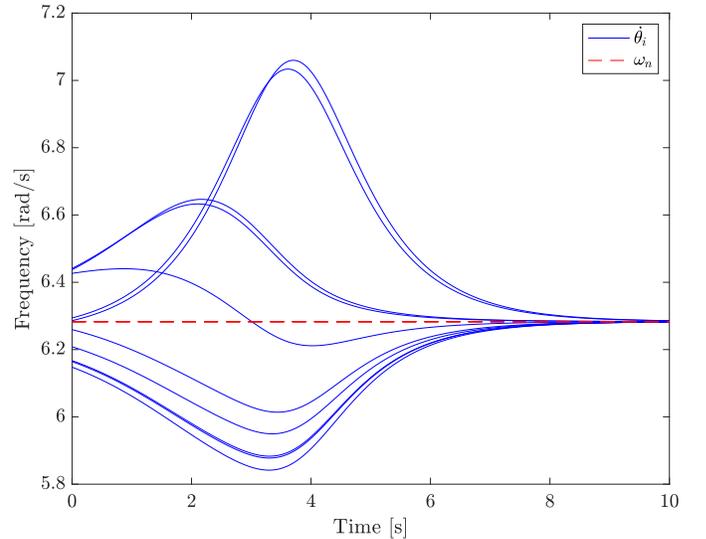}%
\label{frequencies}}
\caption{Evolution of phase, mean phase, and frequencies for a population of oscillators with random initial phase}
\label{KuramotoExample}
\end{figure}

For a population of oscillators with a random distribution of initial phase, the trajectories of the phase, mean phase, and frequency are shown in Fig. \ref{KuramotoExample}. Note that oscillators are basically pulled towards the closest mean phase; we can then think of the mean phase, and its projections every $2\pi$, as beacons that the vehicles are attracted to.

Finally, we combine the behavior of a Kuramoto-driven system and the mapping between desired position and phase we have defined. This combination constitutes the coordinating layer of our algorithm. Through Kuramoto the vehicles agree on a mean phase for the entire network, and because of the definition of the mapping given by Eqs. \ref{grDef} and \ref{gr0}, the vehicles then attempt to cross the intersection exactly when the mean phase is equal to the offset of the road. As such, the synchronizing Kuramoto layer allows vehicles to negotiate the crossing state of all intersections in the network, regardless of their distance to those intersections.

\subsection{Phase, Offset and Wavelength Constraints}

Three different types of constraints need to be satisfied so that the behaviour of oscillators shown in Fig. \ref{KuramotoExample}, corresponds to solving both the intra-junction and the inter-junction problem; these are:

\begin{enumerate}
    \item No two vehicles in the same road segment are being pulled towards the same beacon; this guarantees \textbf{spacing} between vehicles in the same road.
    \item The phase offsets for intersecting roads place the vehicles in the intersection at different times; this guarantees alternate \textbf{servicing} at the intersection.
    \item The phase of a vehicle as it goes from one road segment to the next segment of the same road (i.e., as it goes straight through an intersection, without turning) does not change (in the unit circle); this guarantees \textbf{continuity} of the through flow, thereby reducing energy losses due to re-synchronization.
\end{enumerate}

Vehicles can meet the spacing constraint by properly correcting their phase when a conflict is detected, which mostly occurs when entering a new road segment. Recall that Kuramoto feedback pulls an oscillators towards whichever mean-phase attractor is closer to its current phase. If we define $\Psi_i$ as the projection of the mean-phase closest to the phase $\theta_i$ of vehicle $i$, according to Eq. \eqref{MeanPhaseProjection}, we can write a phase resetting condition for the vehicles that guarantees the spacing constraint:


\begin{equation}\label{resetting}
    \theta_i = \min(\theta_i,\Psi_j-\pi-\epsilon) \quad \forall j \in \{j|s_j>s_i\}
\end{equation}

\noindent where $\epsilon$ is a very small number. By saturating $\theta_i$ in this way, we make sure that $\Psi_i\neq\Psi_j$, which means that no two agents on the same road segment are pulled towards the same attractor. 


\begin{figure}[t]
\centering
\subfloat[Inter-junction diagram]{\includegraphics[width=\columnwidth]{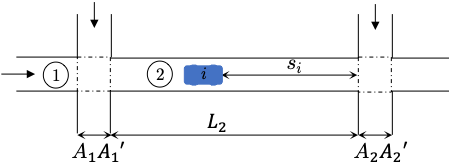}%
\label{sections}}
\hfil
\subfloat[Intra-junction diagram]{\includegraphics[width=\columnwidth]{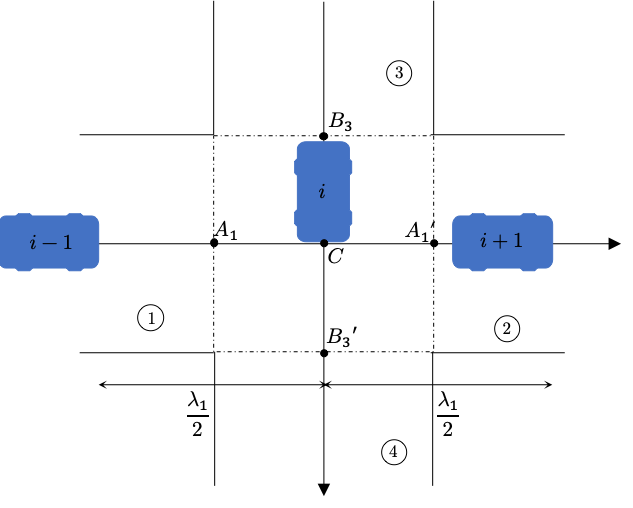}%
\label{intersections}}
\caption{Variable definition as seen within and between junctions}
\label{RoadGeometry}
\end{figure}

To write the safe servicing and continuity constraints mathematically, we consider the variable definitions in Fig. \ref{RoadGeometry}, where we draw a representative intersection zone. Points $A_1$ and $B_3$ correspond to the origins of road segments 1 and 3; that is, the phase at those points is the offset of the respective road segment. Point $C$ represents the intersecting point between the trajectories of vehicles going straight through both roads. Here, and for the rest of the paper we consider an intersection of two one-way roads with only two conflicting traffic movements for the sake of simplicity. More practical traffic scenarios can be accounted for by partitioning the wavelength into however many flows are necessary.

The servicing constraint, which directly relates to solving the intra-junction problem, aims to maximally space out vehicles crossing the intersection from different roads. It is a constraint on the offset of each road that guarantees that each traffic flow is serviced during a different portion of the cycle. Considering the scenario drawn in Fig. \ref{RoadGeometry} we can see that maximal spacing for vehicle $i$ from the vehicles that cross the intersection before and after itself occurs if it reaches the intersection (i.e. point $C$) exactly between them. Now vehicles $i+1$ and $i-1$ are separated by a full wavelength $\lambda_1$, or by $2\pi$ radians in the phase domain (as follows from equation \eqref{grLambda}) . It follows that the distance between vehicle $i$ and $i-1$ should be half a wavelength, or $\pi$ radians in the phase domain. We can show that this is achieved if the mappings of roads 1 and 3 satisfy the following constraint, which relates the phases of point C as mapped by the mappings of each road.

\begin{equation} \label{servicingGeneral}
g_1^{-1}(A_1C)=\pi+g_3^{-1}(B_3C)
\end{equation}

\noindent where $g_p^{-1}(s)$ is the inverse of the mapping \eqref{gr} for road segment $p$. The arguments $A_1C$ and $B_3C$ are the distances between each road's entrance to the intersection and the collision point. For our proposed mapping \eqref{grDef}, the above equation can be rearranged as:

\begin{equation} \label{servicing}
\phi_3-\phi_1=2\pi(\frac{A_1C}{\lambda_1}-\frac{B_3C}{\lambda_3})-\pi
\end{equation}

Finally, the inter-junction coordination problem can be solved automatically by ensuring continuity between mappings as vehicles go from one road segment to the next. That is, we guarantee that the phases at points $A_1$ and $B_3$ are the same when mapped by roads 1 and 2, and roads 3 and 4 respectively. Recalling that points $A_1$ and $B_3$ are the origins of the intersection region, and using equations \eqref{grDef} and \eqref{gr0}, this amounts to:

\begin{equation}
\begin{split}
    g_2^{-1}(L_2+A_1A_1')&=g_1^{-1}(0)=\phi_1\\
    g_4^{-1}(L_4+B_3B_3')&=g_3^{-1}(0)=\phi_3
\end{split}
\end{equation}

Rearranging according to our affine mapping of equation \eqref{gr}, we can express the constraints in terms of the offsets of the roads:

\begin{equation}\label{continuity}
\begin{split}
    \phi_1-\phi_2 &= \frac{2\pi}{\lambda_2}(L_2+\bar{A_1A_1'}) \;(\text{mod } 2\pi)\\
    \phi_3-\phi_4 &= \frac{2\pi}{\lambda_3}(L_4+\bar{B_3B_3'}) \;(\text{mod } 2\pi)\\
\end{split}
\end{equation}

It is worth noting that Eq. \eqref{continuity} can only partially guarantee continuous flow as vehicles travel along a corridor of intersections. For one, the constraint cannot be imposed to turning flows, since the servicing constraint ensures the destination road segment of a turning vehicle will be $\pi$ radians out of phase with respect to its road of origin. Another scenario where flow is disrupted occurs when another vehicle turns into the destination section of the vehicle going straight. In this situation, because of the spacing constraint, the latter vehicle will be forced to slow down to catch the upstream wave. Finally, while the wavelengths can be thought of as adjustable variables in constraint \eqref{continuity}, wavelengths are also constrained by their relationship with frequency and velocity through Eq. \eqref{OmegaConstraint}. Specifically, a change in wavelength from one section to the next would force a change in desired speed through Eq. \eqref{OmegaConstraint} in order to maintain a constant natural frequency, creating an undesirable acceleration or deceleration event. For the rest of this work, we therefore assume that wavelengths and desired speeds are the same across all roads in the network, and we drop the road identifying index $p$ for $\lambda$ and $v^d$.
    
In guaranteeing spacing, safety and continuity to solve the coordination problem at both scales, we have introduced two different types of constraints. The spacing constraint \eqref{resetting} is a constraint on the actual phase of the vehicles; it forces vehicles to push their desired phase back, and with it the time at which it will cross the intersection. This constraint needs to be checked for and implemented continuously, although it will mostly become active when vehicles change road segments. The servicing and continuity constraints, on the other hand, are constraints on the constant design variables of the network, namely the offsets and wavelengths of the roads, and they are chosen before any vehicles enter the network. Along with the desired speed $v^d$, these design variables determine the maximum throughput of the network as we will discuss in subsequent sections.

\subsection{Optimal mean-phase tracking}

So far we have discussed the dynamics of a vehicle's desired phase, which is then mapped to a desired position. In previous work \cite{Rodriguez2018}, we propose a linear feed-forward/feedback tracker that uses this signal as reference. Further insight into the behaviour of the system of coupled oscillators allows us to propose here a more sophisticated tracking approach, namely, a model predictive optimal controller that minimizes the jerk of vehicles using predictions of both the arrival time imposed by the phase dynamics and the behaviour of other vehicles.

We can show that the synchronizing layer described above determines the time $T_i(t)$ at which the vehicle $i$ should ideally arrive at the intersection. Indeed, the computation of $T_i(t)$ follows from the properties of Eq. \eqref{KuramotoMean}, where the mean-phase $\Psi$ oscillates with a constant frequency $\omega_n$ \cite{Strogatz2000}.

\begin{equation}
    \dot{\Psi}(t)=\omega_n(t)
\end{equation}{}

As described in the previous section, vehicle $i$ should reach the intersection when its phase is already tracking its mean phase beacon, which is in turn equal to the offset of the road:

\begin{equation}
    \theta_i(T_i)=\Psi_i(T_i)=\phi_p
\end{equation}

It follows from the previous two equations that for vehicle $i$ at time $t$ the expected time of arrival at the intersection is given by:

\begin{equation}
    T_i(t) = \frac{\phi_p-\Psi_i(t)}{\omega_n}
\end{equation}

Since the vehicle enters the intersection at time $T_i$, in synchrony with its mean phase beacon, its desired position, velocity and acceleration are also known:

\begin{equation}\label{finalTimeCondition}
    \begin{split}
    s_i(t+T_i) &= 0\\
    v_i(t+T_i) &= v^d\\
    a_i(t+T_i) &= 0
    \end{split}
\end{equation}

Assuming vehicles can control their jerk, or their change in acceleration, through accurate lower level powertrain and vehicle dynamics controllers, we model these vehicles as third order dynamical systems. Note that we choose a third order system here, instead of the second order system traditionally used to model vehicles, because it will yield smoother acceleration profiles. With the third order model, the state variables for each vehicle are then: (i) its distance to the intersection, along the path of the road; (ii) its velocity; and (iii) its acceleration. The input is the jerk of the vehicle:

\begin{equation}\label{stateEquations}
    \begin{split}
    \dot{s_i}(t) &= v_i(t)\\
    \dot{v_i}(t) &= a_i(t)\\
    \dot{a_i}(t) &= u_i(t)
    \end{split}
\end{equation}

The control input $u_i(t)$ that places the vehicle at the intersection at the right time, can be the solution of an optimization problem that minimizes mean square jerk:

\begin{equation}\label{optimalControlProblem}
    \min \int_t^{t+T_i(t)}\frac{1}{2}u_i(\tau)^2 d\tau
\end{equation}

Subject to:

\begin{center}
State dynamics \eqref{stateEquations}\\
Terminal time conditions \eqref{finalTimeCondition}
\end{center}
\begin{equation}\label{ineqConstraints}
    \begin{split}
        s_j(t)-s_i(t)-S \leq 0\\
        a_{i,\text{min}}\leq a_i(t) \leq a_{i,\text{max}}\\
        v_{i,\text{min}}\leq v_i(t) \leq v_{i,\text{max}}
    \end{split}
\end{equation}

The additional inequality constraints guarantee that the vehicle stays a safe distance $S$ from its leading vehicle $j$, and that the acceleration and velocity are bounded.

The solution to the problem without the inequality constraints \eqref{ineqConstraints} can be determined analytically by performing a Hamiltonian analysis. This approach is similar to the work of Malikoupoulos et al. in \cite{Rios-torres2017,Zhao2018}, where the solution to a second order dynamical system, where the input is acceleration rather than jerk, is presented. In our case, the optimal trajectories for the input and the states, denoted with an asterisk, are given by:

\begin{equation}\label{unconstrainedSol}
    \begin{split}
     u^*(t) &= -\frac{1}{2}c_1 t^2 +c_2 t-c_3\\
    a^*(t) &= -\frac{1}{6}c_1 t^3 +\frac{1}{2}c_2 t^2 -c_3 t+c_4\\
    v^*(t) &= -\frac{1}{24}c_1 t^4 +\frac{1}{6}c_2 t^3 -\frac{1}{2}c_3 t^2 +c_4 t+c_5\\
    s^*(t) &= -\frac{1}{120}c_1 t^5 +\frac{1}{24}c_2 t^4 -\frac{1}{6}c_3 t^3 +\frac{1}{2}c_4 t^2 +c_5 t+c_6\\ 
    \end{split}
\end{equation}

The constants $c_{1,...,6}$ in the above equations are integration constants, and they can be solved for by imposing initial and final time conditions. The initial conditions are given by the current state of the vehicle at time $t$, and the final conditions are given in equation \eqref{finalTimeCondition}. The resulting system of equations is linear, and it is solved by inverting a 6-by-6 matrix and multiplying it by the concatenated vector of initial and final conditions.

The above is the solution to the unconstrained problem; the solution to the constrained problem can be determined numerically by discretizing and using a quadratic programming solver. This type of optimization is well-understood, convex, and not computationally prohibitive. It can therefore be performed online at every time step when constraints are active. As such, our proposed solution method consists of computing the analytical solution to the unconstrained problem, and checking for constraint activity. If no constraint is infringed upon by the analytic unconstrained solution, we execute the computed input trajectory. Otherwise, we use this candidate solution as the initial guess to the quadratic programming solver and implement the constrained solution instead.

\subsection{Summary}

To summarize the workings of our algorithm, let us recount the actions vehicle $i$ takes at any given time $t$, after it receives the phase and mapping information from the rest of network:

\begin{enumerate}
    \item If the vehicle has just entered a new road segment, it selects its initial phase $\theta_i(t)$ to match its current position according to the mapping of the road.
    \item It computes the order parameter of the system of oscillators (i.e the mean-phase $\Psi$ and coherence $r$ of the network), as well as the projection $\Psi_i$ of $\Psi$ closest to its own phase.
    \item If both its phase and that of its leading vehicle $j$ are most proximal to the same mean-phase beacon ($\Psi_i = \Psi_j$), the vehicle pushes its phase backwards by $\pi + \epsilon$ radians from the beacon tracked by its leader ($\theta_i = \Psi_j -\pi-\epsilon$).
    \item It computes the time it should arrive at the intersection $T_i(t)$ given the current mean-phase.
    \item It determines the optimal trajectory of its state and input that minimizes jerk, according to the analytical solution to the unconstrained optimization problem.
    \item If the solution violates constraints, it solves the constrained optimization problem numerically.
    \item It updates the value of its phase through the Kuramoto equation.
    \item It implements the first input command according to the generated input trajectory.
\end{enumerate}

\begin{figure}[!h]
    \centering
    \includegraphics[width=\columnwidth]{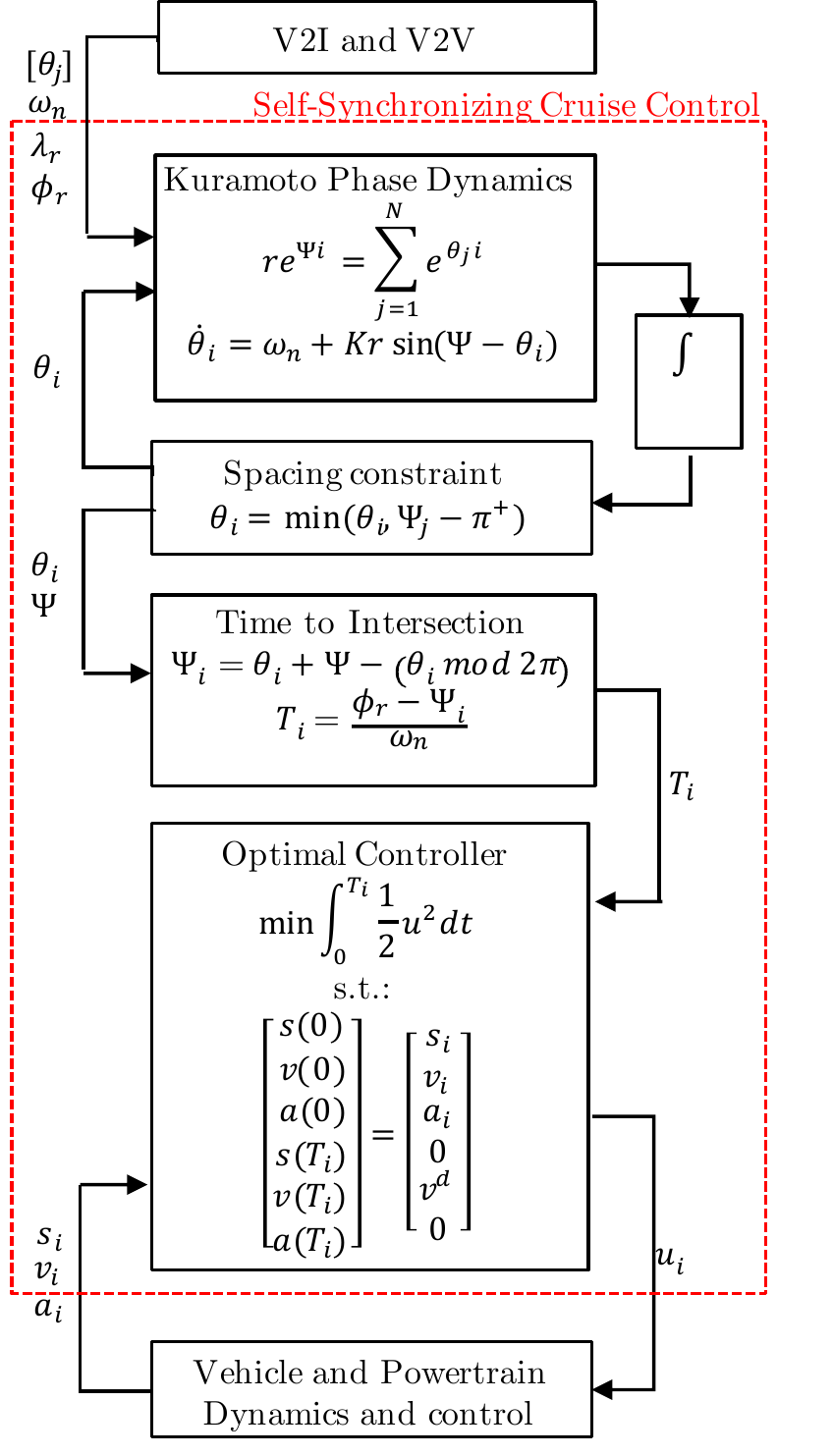}
    \caption{Control Architecture}
    \label{controlArch}
\end{figure}

Fig. \ref{controlArch} summarizes this process in a block diagram. The result of following this protocol is that vehicles cross the intersection at different times, and that acceleration maneuvers as they go from one intersection to next are not very aggressive.

\section{Designing for Traffic Flow and Safety}
Before looking at the performance of our strategy in simulation, we can discuss some of its anticipated implications in terms of traffic flow and density. By virtue of Kuramoto synchronization, all vehicles oscillate at the same frequency once coherence is achieved. In fact, in our current formulation, this frequency corresponds to the natural frequency we choose for the network:

\begin{equation}
    \omega_n = 2\pi \frac{v^d}{\lambda}
\end{equation}

The flow of vehicles in each road is directly related to this frequency given that vehicular flow is the product of velocity and density. Maximum density is nothing more than the inverse of the wavelength, because in each road there can only be one vehicle per wave, and vehicles are spaced by one wavelength (or more). The maximum possible flow is then, in its traditional units of vehicles per hour:

\begin{equation}
    q = \frac{v^d}{\lambda} (3600)=\frac{\omega_n}{2\pi}(3600)
\end{equation}

We can then expect that for input flows below the selected natural frequency, the algorithm will be able to meet the traffic demand. For higher input flows, a queue will start to form at the entrances of the network as vehicles wait to track non-occupied wave crests. 

Velocity and wavelength should be chosen to produce a natural frequency higher than the demand of the road. However, this is not the only constraint on these two variables, since the spacing of vehicles as they cross the intersection also depends on these variables. In fact, from analysing Fig. \ref{RoadGeometry}, we can determine that the gap in seconds between a vehicle at the collision point of the intersecting paths and the vehicle that just crossed is given by:

\begin{equation}
    g = \frac{1}{v^d}(\frac{\lambda}{2}-S)
\end{equation}

Where S is a safety distance that needs to be larger than the occupied portion of the wave, that is, the length plus the width of the vehicles.

\begin{figure}[t]
    \centering
    \includegraphics[width=\columnwidth]{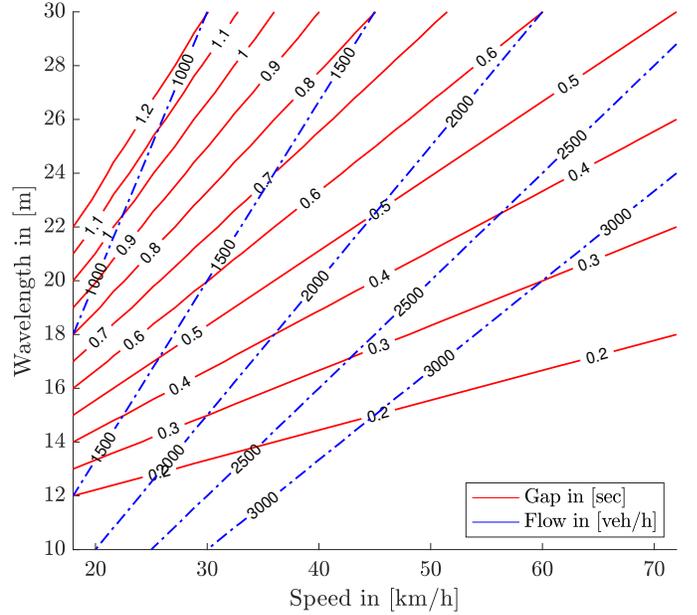}
    \caption{Relationship between speed, wavelength, flow and safety}
    \label{FlowGap}
\end{figure}

Having defined the relationship between our design variables $v^d$ and $\lambda$, we can look at the inherent trade-offs between increasing the maximum throughput of the network and maintaining enough spacing between vehicles at the intersection. Fig. \ref{FlowGap} shows this trade-off in our design space. We have drawn lines of constant throughput and lines of constant safety gap. We can see that to increase the safety gap, one might decide to choose larger wavelengths; however, since this will reduce the density of the roads, throughput will be affected. Alternatively, if one wishes to increase throughput, the simplest way to ``cross'' dashed-blue lines is to increase speed, but this comes at the cost of reducing the safety gap. Another inherent trade-off that does not show up in Fig. \ref{FlowGap}, as it is more difficult to compute analytically, is the energy/fuel cost associated with having longer wavelengths. If a vehicle enters a road completely out of phase, the acceleration/deceleration maneuver it will need to perform is larger in roads with larger wavelengths. This translates to a higher kinetic energy change, and with it, some potential waste of fuel.

\section{Simulation Studies}

\begin{figure}[!h]
    \centering
    \includegraphics[width=\columnwidth]{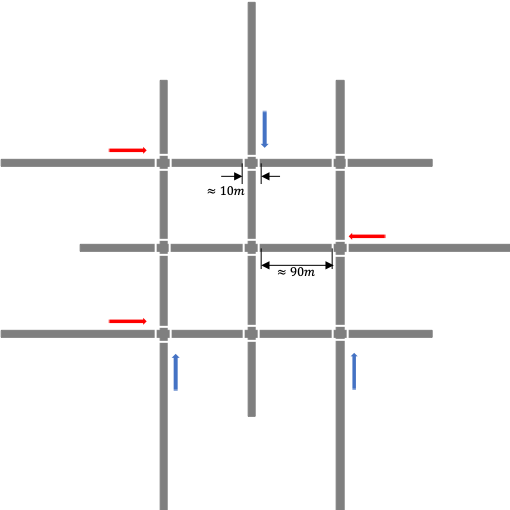}
    \caption{Network of 9 intersections used in simulation. }
    \label{networkDiagram}
\end{figure}

In this section, we study the performance and characteristics of our proposed strategy in simulation. We consider a network of one-way roads consisting of 9 intersections and 24 road segments where  vehicles can either go straight or turn right; a snapshot of this network is shown in Fig. \ref{networkDiagram}. The road segments are approximately 90 meters long, and the straight segment of the intersections is approximately 10 meters long. The entry roads to the network are assumed to be longer, at 200 meters. For this network, where the origins of the mappings between connected road segments are 100 meters apart, a 20 meter wavelength would satisfy the continuity and servicing constraints of Eqs. \eqref{continuity} and \eqref{servicing} if we choose offsets of $0$ and $\pi$ for horizontal and vertical roads respectively.

\subsection{State Trajectories}

\begin{figure}[!h]
    \centering
    \includegraphics[width=\columnwidth]{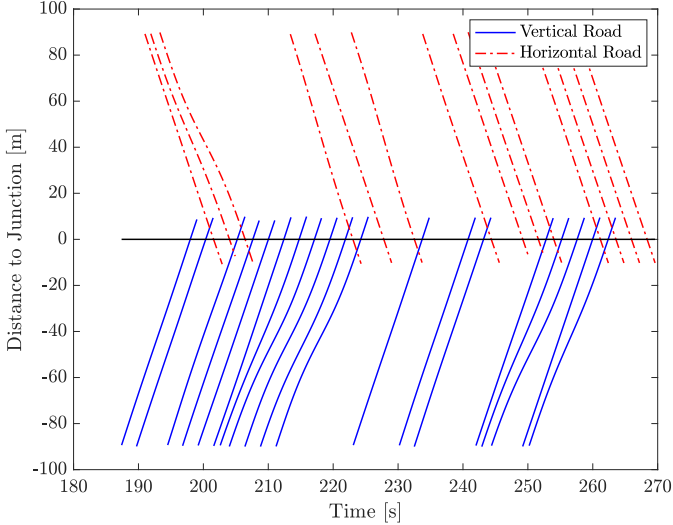}
    \caption{Example position trajectories for a group of vehicles approaching the same intersection along the horizontal (red dotted-solid line) and vertical (blue solid line) roads.}
    \label{posTrajectoriesJunction}
\end{figure}

Fig. \ref{posTrajectoriesJunction} shows the distance to the intersection as a function of time for a group of vehicles approaching the intersection at the center of the network. In this figure, as in subsequent ones, the color of the curve indicates whether the vehicle is travelling down a horizontal (dashed red) or a vertical (solid 
blue) road segment, and, for clarity, we flip the sign of the distance along the horizontal directions. We can see that red and blue lines cross the 0 line at different points, meaning that vehicles enter the intersection at different times. Moreover, the plot illustrates how vehicles space out evenly along the same road.

\begin{figure}[!h]
    \centering
    \includegraphics[width=\columnwidth]{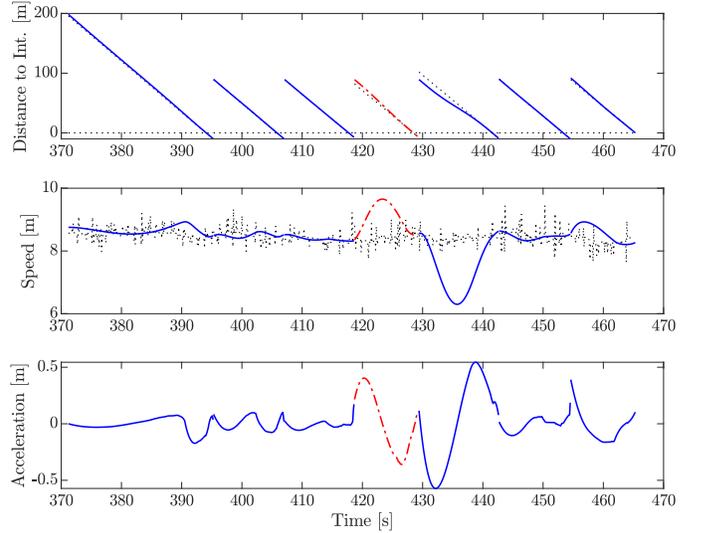}
    \caption{Example position, velocity and acceleration trajectories for a single vehicle travelling through the network in horizontal (red solid-dotted) or vertical roads (solid blue), along with the reference mean phase and frequency (dotted black).}
    \label{StateTrajectories}
\end{figure}

We can also look at the position, velocity and acceleration of a single vehicle as it travels through the network, which we show in Fig. \ref{StateTrajectories}. Here, we have also plotted in solid blue the vertical segments, and in dashed red the horizontal ones. We can see that as the vehicle goes straight through the intersections its velocity profile stays relatively flat, as promoted by the continuity constraint we impose on the mapping and the fact the consensus occurs at a network level. When the vehicle turns in the third intersection it needs to adjust its speed to match the offset of the new road it travels on. The same thing happens as it turns right again in the next intersection.

\subsection{Fuel Consumption and Delay Time Results}

\begin{figure}[!h]
    \centering
    \includegraphics[width=\columnwidth]{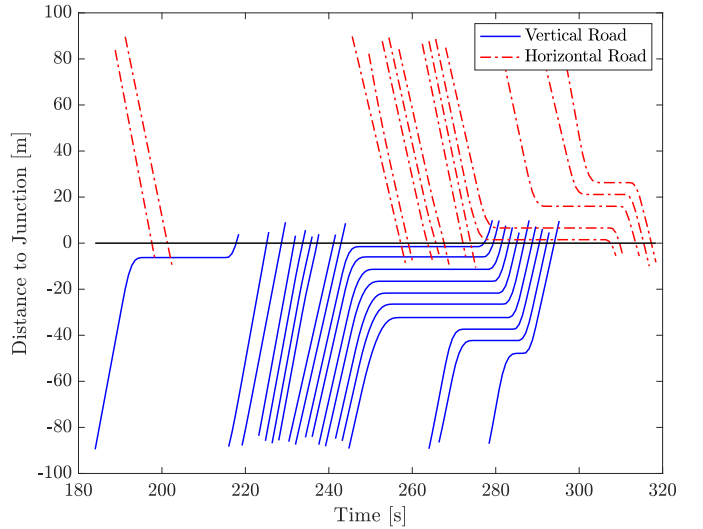}
    \caption{Example baseline position trajectories for vehicles approaching an intersection controlled by a traffic light.}
    \label{posTrajectoriesJunctionBaseline}
\end{figure}

We can evaluate the fuel consumption and delay time of vehicles using our strategy compared to simulated human drivers controlled by traffic lights. The baseline drivers are governed by a modified Gipps car following model \cite{Gipps1981} as implemented in Aimsun, an established traffic simulator. We set the input flow of all entry roads at 750 vehicles per hour, with a turn percentage of 20\%. The arrival process of vehicles into the network is the main source of stochasticity in our simulation, and it  is modeled as a Poisson arrival process, as is traditionally done in traffic simulation \cite{barcelo2010}. We choose a traffic light cycle of 60 seconds, with 25 seconds of green time for each flow and 10 seconds of clearing time. Furthermore we offset the green time of the lights in pursuit of the "green wave" effect, which occurs when vehicles catch several green windows in a row as they travel down an arterial corridor. We run the baseline simulation for 10 minutes of simulated time, and we replicate the scenario with the same vehicle injection times and paths, but using our Kuramoto strategy instead. Fig. \ref{posTrajectoriesJunctionBaseline} shows the baseline position trajectories corresponding to the same vehicles shown in Fig. \ref{posTrajectoriesJunction} in the last sub-section.

The described simulation consists of about 750 vehicles, but for our comparisons we consider only the 100$^{th}$ through 600$^{th}$ vehicles. In this way, we allow for the network to build some capacity, and we don't consider vehicles who don't finish their path before the simulation is stopped.

\begin{figure}[b]
    \centering
    \includegraphics[width=\columnwidth]{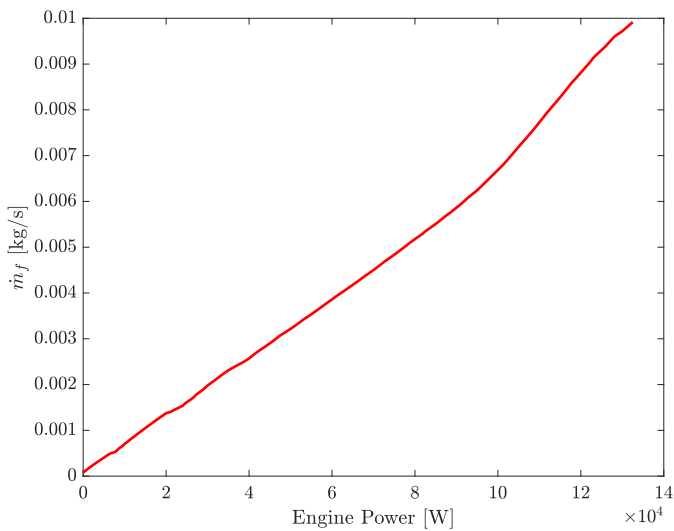}
    \caption{Optimal fuel rate vs. engine power for a 1.2 liter gasoline engine}
    \label{optimalFuel}
\end{figure}

We are interested in looking at two metrics relevant for traffic performance evaluation: fuel consumption and delay time. The delay time is simply the difference between a vehicle's travel time and its corresponding minimum travel time had it cruised at the desired speed of the road, normalized by the total distance traveled. To calculate fuel consumption, we use a fuel map for a 1.2 liter gasoline engine. The map translates every engine torque and speed pair to a fuel rate $\dot{m_f}$. To use it, we first calculate the wheel power required to meet the acceleration and velocity trajectories imposed by the driver. We then estimate the corresponding engine power by assuming an efficiency ratio for the transmission. Finally, we say the vehicle uses the minimum fuel rate associated with this engine power demand, which assumes the transmission can operate at the required engine torque and speed combination. Fig. \ref{optimalFuel} shows the optimal fuel rate vs. engine power line we get.

\begin{figure}[t]
    \centering
    \includegraphics[width=\columnwidth]{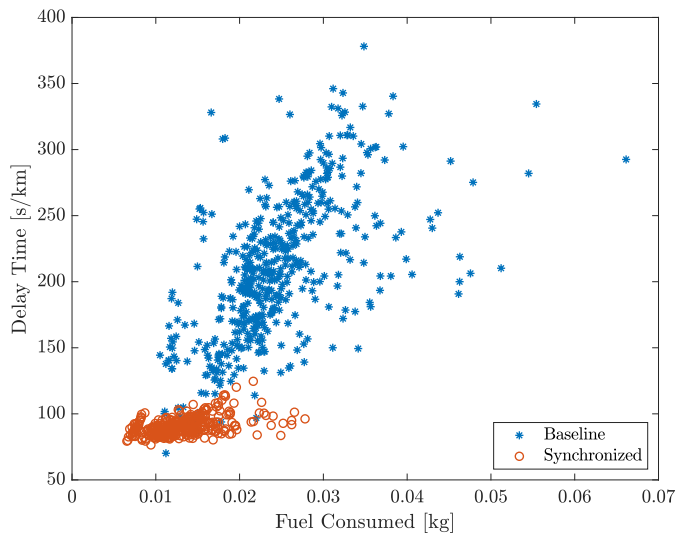}
    \caption{Delay time and fuel consumed for each vehicle in the simulation of the baseline (green) and proposed strategy (blue) }
    \label{FuelAndDelay}
\end{figure}

Fig. \ref{FuelAndDelay} shows the delay time and fuel consumed by each of the vehicles for both the baseline and proposed scenario. When we compare the average of both point clouds, we find that our proposed strategy leads to a 48\% and 57\% reduction in fuel consumption and delay time respectively for this particular scenario. Furthermore, we note a significant reduction in the spread of the point cloud, meaning that there is less variability in the anticipated behaviour of the vehicles. Indeed, in the baseline, a vehicle that encounters a desirable green wave of traffic light sequences can traverse the network quickly without stop-and-go behaviour, whereas vehicles that are less lucky are forced to stop at several intersections in sequence. 

\begin{figure}[b]
    \centering
    \includegraphics[width=\columnwidth]{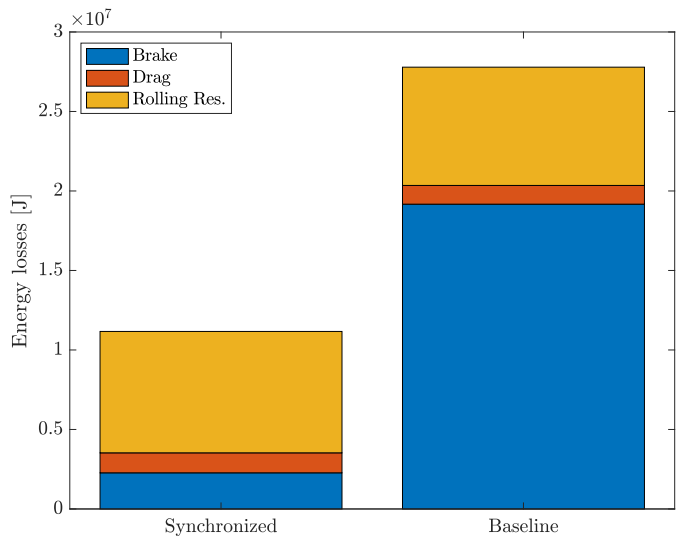}
    \caption{Energy losses by type (brake, drag, and rolling resistance) for the synchronization strategy and the baseline strategy }
    \label{EnergyExp}
\end{figure}

If we compute the work done by negative propulsive forces (i.e. braking), drag forces, and rolling resistance forces in our model for longitudinal vehicle dynamics, we can see where energy losses are incurred. The savings in fuel consumption can then indeed be attributed to a reduction in energy losses due to braking. In other words, our strategy improves performance by reducing stop-and-go behaviour, as expected. Fig. \ref{EnergyExp} shows the result of this energy balance.

\section{Conclusion}
In this paper we present a control strategy for connected and autonomous vehicles that solves the intersection coordination problem in both of its scales. That is, our strategy synchronizes vehicles crossing the same intersection, and it smooths the flow from one intersection to the next. This is achieved by defining a mapping between a vehicle's position and its corresponding phase in a virtual system of oscillating agents coupled by the Kuramoto equation. The mapping, with its safety constraints within the intersection and continuity constraints between intersections, guarantees the desired behaviour of the reference position. This reference is then the tracked through an optimal control problem that is first solved analytically, and then numerically if constraints are violated. The resulting strategy saves both fuel and travel time, and reduces the variability in these metrics seen across the fleet.

Our work here shows the potential of using self-synchronizing Kuramoto consensus to coordinate CAVs. Future work can extend the approach to include traffic lights and mixed traffic (i.e. human-driven and autonomous). Starting from the algorithm presented here, it is quite straightforward to extract signal phase and timing information and display it as a traffic light. The traffic light, in turn, can control human driven vehicles. One can then study how the mixed traffic fleet affects the performance of the network.

\section*{Acknowledgment}
The information, data, or work presented herein was funded in part by the Advanced Research Projects Agency-Energy (ARPA-E), U.S. Department of Energy, under Award Number DE-AR-0000801. The authors gratefully acknowledge this support.

\ifCLASSOPTIONcaptionsoff
  \newpage
\fi



%
\bibliographystyle{IEEEtran}
\bibliography{IntersectionManagement.bib}

%

%


\begin{IEEEbiographynophoto}{Manuel Rodriguez}
received his B. Sc. degree in engineering sciences, and M.s. degree in mechanical engineering from The Pennsylvania State University, in State College, PA, in 2016 and 2018 respectively. He is currently pursuing his doctoral studies in mechanical engineering at the University of Maryland, College Park, MD. His research interests include optimal control of multi-agent systems.  
\end{IEEEbiographynophoto}


\begin{IEEEbiographynophoto}{Hosam K. Fathy}
received the B.Sc. degree in mechanical engineering from The American University in Cairo, New Cairo, Egypt, in 1997, the M.S. degree in mechanical engineering from Kansas State University, Manhattan, KS, USA, in 1999, and the Ph.D. degree in mechanical engineering from the University of Michigan, Ann Arbor, MI, USA, in 2003. He is currently a professor of mechanical engineering at the University of Maryland. His research interests include reduced-order modeling and optimal control of energy storage and management systems. He was the recipient of the 2014 NSF CAREER award.
\end{IEEEbiographynophoto}




\end{document}